# Inferring CT perfusion parameters and uncertainties using a Bayesian approach

Authors:

Tao Sun, Paul C. Lauterbur Research Center for Biomedical Imaging, Shenzhen Institute of Advanced Technology, China

Roger Fulton, Faculty of Medicine and Health and School of Physics, University of Sydney; Department of Medical Physics, Westmead Hospital, Australia

Zhanli Hu, Paul C. Lauterbur Research Center for Biomedical Imaging, Shenzhen Institute of Advanced Technology, China

Christina Sutiono, Radiology Department, Western Sydney Local Health District, Westmead Hospital, Australia

Dong Liang, Paul C. Lauterbur Research Center for Biomedical Imaging, Shenzhen Institute of Advanced Technology, China

Hairong Zheng, Paul C. Lauterbur Research Center for Biomedical Imaging, Shenzhen Institute of Advanced Technology, China

Corresponding author:
Tao Sun, E-mail: tao.sun@siat.ac.cn   Phone 86-13820733456
Address: 1068 Xueyuan Avenue, Nanshan District, Shenzhen, Guangdong, China

Author contribution:
(I) Conception and design: Tao Sun, Roger Fulton
(II) Administrative support: Tao Sun, Zhanli Hu, Roger Fulton, Dong Liang, Hairong Zheng
(III) Provision of study materials or patients: Tao Sun, Roger Fulton, Christina Sutiono
(IV) Collection and assembly of data: Tao Sun, Roger Fulton, Christina Sutiono
(V) Data analysis and interpretation: Tao Sun, Zhanli Hu, Roger Fulton
(VI) Manuscript writing: All authors
(VII) Final approval of manuscript: All authors

ABSTRACT

*Background:* Computed tomography perfusion imaging is commonly used for the rapid assessment of patients presenting with symptoms of acute stroke. Maps of perfusion parameters, such as cerebral blood volume (CBV), cerebral blood flow (CBF), and mean transit time (MTT) derived from the perfusion scan data, provide crucial information for stroke diagnosis and treatment decisions. Most CT scanners use singular value decomposition (SVD)-based methods to calculate these parameters. However, some known problems are associated with conventional methods. *Methods:* In this work, we propose a Bayesian inference algorithm, which can derive both the perfusion parameters and their uncertainties. We apply the variational technique to the inference, which then becomes an expectation-maximization problem. The probability distribution (with Gaussian mean and variance) of each estimated parameter can be obtained, and the coefficient of variation is used to indicate the uncertainty. We perform evaluations using both simulations and patient studies. *Results:* In a simulation, we show that the proposed method has much less bias than conventional methods. Then, in separate simulations, we apply the proposed method to evaluate the impacts of various scan conditions, i.e., with different frame intervals, truncated measurement, or motion, on the parameter estimate. In one patient study, the method produced CBF and MTT maps indicating an ischemic lesion consistent with the radiologist's report. In a second patient study affected by patient movement, we showed the feasibility of applying the proposed method to motion corrected data. *Conclusions:* The proposed method can be used to evaluate confidence in parameter estimation and the scan protocol design. More clinical evaluation is required to fully test the proposed method.

**Keywords:** stroke, CT perfusion, Bayesian inference.

I. INTRODUCTION

Stroke is the third leading cause of death in the US, killing about 140000 Americans each year (1). There are two types of strokes, ischemic and hemorrhagic, for which the medical management can be quite different. About 87% of all cases are ischemic strokes (2), meaning there is a complete or partial blockage of the blood supply from a cerebral artery. Standard treatment options for ischemic strokes are to either dissolve or remove the blood clot (i.e., thrombolysis or thrombectomy). Early treatment decisions are crucial for the survival and recovery of stroke patients. To aid the diagnosis and selection of the appropriate treatment, the location and volume of the infarct core and penumbra are often calculated—the infarct core indicates the brain region that is not salvageable, and the penumbra indicates the brain region that is at risk of progression to infarction. The penumbra usually surrounds the ischemic core and may still be salvageable. Classification of the infarct core and penumbra is often derived by thresholding certain brain hemodynamic parameters, e.g., cerebral blood volume (CBV, mL/100g), cerebral blood flow (CBF, mL/100g/min), mean transit time (MTT, s), and time to peak (TTP, s). For example, in (3), the ischemic core was defined as regions with relative MTT $\geq$ 145% of the normal tissue and absolute CBV < 2.0 ml/100 g.

Computed tomography perfusion and dynamic contrast-enhanced magnetic resonance imaging are the primary imaging techniques used to obtain the hemodynamic parameters of the brain. In this study we are interested in CT perfusion, a procedure performed after the administration of intravenous iodine contrast to monitor the first pass of a contrast bolus through the cerebral circulation. Based on the well-known indicator dilution theory (4), from a CT perfusion scan one can obtain the hemodynamic functional parameters related to the blood passage in the tissue, including CBF, CBV, MTT, and TTP. Various approaches, including singular value decomposition (SVD)-based methods, have been applied to solve this inverse problem. In the clinical setting, perfusion analysis is usually done with SVD-based methods, the implementation of which varies across vendors.

Unfortunately, SVD-based methods have some known problems, which make their derived hemodynamic parameters unreliable. First, they make assumptions about the ideal underlying physiological model. For example, the residual function is assumed to be an impulse function, which is not valid for a real scan. Because of this, the estimated parameters can be biased. As shown in (5,6), the SVD-based method can produce biased CBF and MTT estimates. Several authors have proposed advanced deconvolution methods, such as Tikhonov regularization, to address this problem (7–9). Second, these methods often cannot well tolerate imperfect measurements, including relatively long scan intervals (10,11) and truncated measurements in time (12,13). Third, these methods are deterministic, meaning they provide an estimate of the mean of the parameter of interest but not its probability distribution. Factors that affect the reliability of the measurement in a CT perfusion scan, such as the protocol design and perturbance of data, therefore cannot be accounted for in the estimation. For example, a too-long scan interval reduces the available samples for estimation and hence the accuracy. In other cases, patient movement could produce inconsistent voxel attenuation measurements that would inevitably increase the uncertainty of the estimation. Post-processing (e.g., spatial and temporal filtering) may introduce further uncertainty of estimation. For quantification, it is desirable to account for such randomness with proper probability theory (14). Therefore, a probabilistic inference method seems warranted to infer both the hemodynamic parameters and their associated uncertainties.

In this paper, we propose such an accurate and probabilistic method and show its potential value in clinical application. In Section II, we formulate the estimation problem as a Bayesian inference problem. A variational technique, mean-field approximation, is applied, which makes the problem more tractable as a classic expectation-maximization problem. The posterior distributions (with mean estimate and variance) of CBF and MTT can be obtained. Their coefficients of variation (CoV) are used to indicate the uncertainty in inference, which reflects the reliability of a given measurement. In Section III, we first demonstrate the superiority of the proposed method over the conventional method in simulation studies where ground truth is available. We also demonstrate the potential usage of the proposed method in evaluating different suboptimal scan conditions. Finally, patient scans were processed with the proposed method. The results are discussed in Section IV.

## II. MATERIALS AND METHODS

We first present the implementation of the SVD-based methods. Then the forward maximum log-likelihood model of CT perfusion is presented. We then detail the proposed Bayesian inference algorithm and the quantitative analysis design.

### A. SVD-based methods

In a CT perfusion scan, a mask image is first acquired. The enhanced image at a single time point can be obtained by subtracting the mask image from an image acquired later in the scan. The enhancement of tissue depends on the iodine concentration, which at one voxel can be represented as:

$$g(t) = h(t) * C_a(t), \quad (1)$$

where $h(t)$ is the flow-scaled residual function, indicating the fraction of contrast that remains in the voxel at time $t$ after its arrival; $C_a(t)$ is the arterial input function (AIF), indicating the contrast change in the arteries; and $*$ represents the convolution operation. Equation (1) describes the well-known indicator dilution theory

((4), Fig. 1). Based on this theory, one can obtain $h(t)$ from a CT perfusion scan, which can be used to derive the hemodynamic functional parameters related to the blood passage in the tissue, including CBF, CBV, MTT, and TTP which are defined as:

$$\text{CBF} = \frac{1}{\rho} \cdot \max(h(t)),$$
$$\text{CBV} = \frac{1}{\rho} \int_0^\infty h(\tau) d\tau,$$
$$\text{MTT} = \text{CBV}/\text{CBF}, \qquad (2)$$
$$\text{TTP} = \underset{t}{\text{argmax}} \, g(t)$$

where $\rho$ is the mean density in the capillary bed, which is set to 1.04 g/ml in this study (15). A detailed explanation of how (2) is derived can be found in (16).

A straightforward way to obtain the residual function h(t) is to perform deconvolution on (1) [7]. In practice, g(t) is sampled at discrete time points. Thus, the deconvolution problem is an inverse algebraic problem, which can be solved by SVD. The deconvolution is performed as illustrated in (17). To suppress noise in the resulting residual function, the least significant eigenvectors are consecutively removed until $h(t)$ has an oscillation index below a certain threshold. In this study,

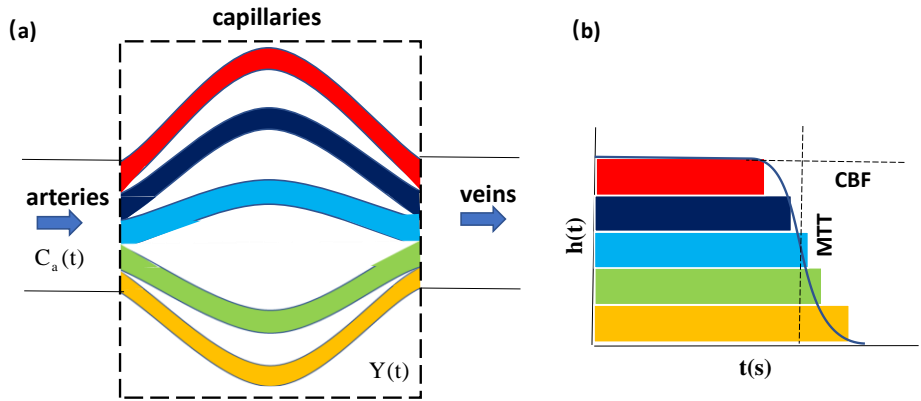

CBF-cerebral blood flow, MTT-mean transit time a threshold of 0.095 is used as suggested in (17). To further suppress the spatial noise, an alternate deconvolution method with Tikhonov regularization is also implemented using the Regularization Tools package (18). The parameter selection and regularization setup are as suggested in (7).

Fig. 1 The indicator dilution theory model (a) describes the blood supply to the tissue at one voxel in the brain. After contrast injection, the bolus enters the voxel via an arterial inlet $C_a(t)$, is diluted in the capillary bed, and leaves the voxel via the venous outlet. Due to the limited spatial resolution of the CT image, the measured concentration Y(t) contain numerous capillary beds as well as arterioles and venules. (b) is he residual function h(t) describes the hemodynamic functions in the physiological model. CBF, CBV, and MTT can be derived from h(t).

## B. Forward maximum log-likelihood model

Starting from (1), the measurement process in a CT perfusion scan can be represented as

$$Y(t) = g(t) + \varepsilon(t)$$
$$= h(\mu, t) * C_a(t) + \varepsilon(t) \quad (3)$$

where $Y(t)$ is the enhanced tissue time curve at the voxel of interest, $\mu$ represents the hemodynamic parameters to be estimated, $h(\mu, t)$ is the flow-scaled residual function, and $\varepsilon(t)$ is the noise. The flow-scaled residual function $h(\mu, t)$ can be further written as a function of the CBF, tissue density $\rho$, and residual function $r(t)$:

$$h(\mu, t) = \text{CBF} \cdot \rho \cdot r(t)$$

According to (19 – 21), $r(t)$ can be modelled as a box shaped function with exponential decay. Hence, $h(\mu, t)$ becomes:

$$\begin{cases} h(\mu, t) = \text{CBF} \cdot \rho \cdot \exp(-\frac{(t-T_0)}{\text{MTT}-T_0}), t \geq T_0 \\ h(\mu, t) = \text{CBF} \cdot \rho, t < T_0 \end{cases}, \quad (4)$$

where $T_0 = 0.632 \times \text{MTT}$. In this study, we consider $\mu$ to contain two parameters, CBF and MTT. Other parameters can be calculated; for example, CBV can be calculated as CBV = CBF × MTT, and TTP can be derived directly from the time-attenuation curve (TAC) as the time when the peak enhancement is reached, as shown in (2). There is no bolus arrival delay modelled here. The convolution in (3) can be expanded by sampling the integral of the arterial enhancement curve:

$$Y(t) = \int_0^t C_a(\tau) \cdot h(\mu, t - \tau) d\tau + \varepsilon(t) \quad (5)$$

As explained in the introduction, $Y(t)$ in practice is sampled at $N$ discrete time points. Without knowledge of the measurement errors, $\varepsilon$ is then modelled as a zero-mean Gaussian distributed in each frame. Since the noise in each frame is temporally independent, we can express the discrete version of (5) as:

$$y_i = g_i(\mu) + \varepsilon_i$$
$$\varepsilon_i \sim N_i(0, \omega_i s^2),\quad (6)$$

where the index of the time frame is $i$, and $\omega_i$ is proportional to the number of the detected photon counts (duration of the $i$th frame as a surrogate). $s^2$ is the variance at a given frame. From (6) we can build the data mismatch term, and the inference problem becomes one of finding the estimate that maximizes the log-likelihood

$$P(Y|\mu) \propto \prod_{i=1}^{N} \exp\left(-\frac{[y_i - g_i(\mu)]^2}{2\omega_i s^2}\right)$$
$$\log P(Y|\mu) = \sum_i^N -\frac{1}{2}\log \omega_i s^2 - \frac{1}{2}(y_i - g_i(\mu))^T \frac{1}{\omega_i s^2}(y_i - g_i(\mu)) + C_0 \quad (7)$$

$$\hat{\mu} = \underset{\mu}{\mathrm{argmax}}\log P(Y|\mu)$$

where $C_0$ is a constant term. In the case where the length of each frame is the same, one can set $\omega_i = 1$. The above optimization problem is essentially equivalent to maximum likelihood estimation, which can be solved by nonlinear least squares fitting, and the way to perform the inference is similar to that of previous perfusion estimation methods (22).

## C. Bayesian inference algorithm

Now we extend the above maximum likelihood estimation to a full Bayesian inference, which provides the probability distribution rather than just the maximum likelihood estimate of a given parameter in (7). The derived distribution will inherently contain the measurement uncertainty of each parameter. According to Bayes' theorem, the posterior of the estimate $\mu$ over measurement $Y$ is:

$$P(\mu|Y) = \frac{P(Y|\mu) \cdot P(\mu)}{P(Y)}, \quad (8)$$

where $P(Y|\mu)$ is the likelihood in (7), and $P(\mu)$ is prior. The fundamental difference here from (7) is that we are no longer interested in the maximum likelihood estimate but the full distribution of $\mu$. Solving an inference problem within the Bayesian framework involves maximizing the posterior probability $P(\mu/Y)$ of $\mu$ given the observed data $Y$. In practice, computing the posterior $P(\mu|Y)$ in a closed form is often intractable. An approximation is therefore often used, for which the Markov chain Monte Carlo (MCMC) and variational Bayesian are two popular choices. To achieve a reasonable computational load, the approximate but efficient variational Bayesian method is used here. The basic idea is to find a simple analytical distribution $q(\mu|Y)$ to approximate the intractable posterior $P(\mu|Y)$ such that the Kullback–Leibler (KL) divergence of these two is minimized. Given the probabilistic model (8), the log evidence can be written as:

$$\begin{aligned}
\log p(Y) &= \int q(\mu|Y) \log p(Y) d\mu \\
&= \int q(\mu|Y) \log \frac{P(Y|\mu)P(\mu)}{P(\mu|Y)} d\mu \\
&= \int q(\mu|Y) \log \frac{P(Y,\mu)q(\mu|Y)}{P(\mu|Y)q(\mu|Y)} d\mu \\
&= \int q(\mu|Y) \log \frac{q(\mu|Y)}{P(\mu|Y)} d\mu + \int q(\mu|Y) \log \frac{P(Y,\mu)}{q(\mu|Y)} d\mu \\
&= E^* \left[ \log \frac{q(\mu|Y)}{P(\mu|Y)} \right] + E^* \left[ \log \frac{P(Y,\mu)}{q(\mu|Y)} \right] \\
&= KL + ELBO
\end{aligned}$$

where $P(Y, \mu)$ is the joint probability, $E^*$ denotes the expectation with respect to $q(\mu|Y)$, ELBO is the lower bound of the evidence, and KL is the divergence between approximate $q(\mu|Y)$ and $P(\mu/Y)$. Because the KL divergence is always positive, ELBO provides a lower bound on the log-likelihood evidence. Therefore, one can maximize ELBO by finding the distribution $q(\mu)$ that approximates the posterior close to the true posterior (or minimizes KL):

$$\hat{q}(\mu) = \underset{q(\mu)}{\mathrm{argmax}} \mathrm{ELBO} = \underset{q(\mu)}{\mathrm{argmin}} E^* \left[ \frac{\log q(\mu|Y)}{\log P(Y,\mu)} \right] \quad (9)$$

Now we have a second KL divergence in the square brackets, which is always positive, and the optimal solution can be found by equating the numerator and denominator.

Considering the **numerator**, we first need to ensure it is tractable. Variational Bayesian factorizes $q(\mu|Y)$ (in physics this is known as mean-field approximation) to sort the parameters into separate groups, each with their own approximate posterior (23). Note that this approximation does not imply that the parameters are uncorrelated for a given measurement. A simple choice of the distribution of the approximated true posterior is multivariate Gaussian. The mean-field approximation theory assumes that the posterior distribution of each parameter is Gaussian. Supposing that we have a number, $k$, of parameters to be estimated, the approximate posterior has the following analytical expression:

$$\begin{aligned}
q(\mu|Y) &= \prod_k N(\mu_k, m_k, \sigma_k) \\
\log q(\mu|Y) &= \sum_k -\frac{1}{2} \mu_k \frac{1}{\sigma_k} \mu_k + \mu_k \frac{1}{\sigma_k} m_k + C_1 \quad (10) \\
&= -\frac{1}{2} \mu^T \sigma \mu + \mu^T \sigma m + C_1
\end{aligned}$$

where $\sigma$ is a matrix with $1/\sigma_k$ as the diagonal elements, $m_k$ is the mean estimate of each parameter, and $C_1$ is a constant term. Then, for the **denominator**, we can insert the likelihood from (7). As for the prior, we choose the conjugate prior—the prior is said to be conjugate to the likelihood if and only if the factorized posterior has the same parametric form (24). Therefore, the conjugate prior has a multivariable Gaussian distribution with mean $m_0$ and variance $\sigma_0^2$. The numerator thus becomes:

$$logP(Y,\mu) = logP(Y|\mu)P(\mu)$$
$$= \sum_i \left[-\frac{1}{2}(y_i - g_i(\mu))^T \frac{1}{\omega_i s^2}(y_i - g_i(\mu))\right] - \frac{1}{2}(\mu - m_0)^T \frac{1}{\sigma_0}(\mu - m_0) + C_2 \quad (11)$$

where $C_2$ is a constant term. To ensure tractability and to allow the method to generalize to any nonlinear model, $g(\mu)$ is approximated by a first-order Taylor expansion about the mode (mean) of the posterior:

$$g_i(\mu) \approx g_i(m) + J_i(\mu - m)$$
$$(J)_{j,k} = \left.\frac{d(g(\mu)_j)}{d\mu_k}\right|_{\mu=m}$$

where J is the Jacobian. With such linearization applied to $g(\mu)$, (11) becomes:

$$logP(Y,\mu)$$
$$\approx \sum_i \left[-\frac{1}{2}(y_i - g_i(m) + J_i(\mu - m))^T \frac{1}{\omega_i s^2}(y_i - g_i(m) + J_i(\mu - m))\right] - \quad (12)$$
$$\frac{1}{2}(\mu - m_0)^T \frac{1}{\sigma_0}(\mu - m_0)$$

By equalizing (12) and (10), we have updated equations for the mean $m$ and $\sigma$:

$$\sigma_{new} = \sigma_0 + \sum_i \frac{1}{\omega_i s} J_i^T J_i$$
$$m_{new} = \frac{\sigma_0 m_0 + \sum_i \frac{1}{\omega_i s} J_i(y_i - g_i(m) + J_i m)}{\sigma_0 + \sum_i \frac{1}{\omega_i s} J_i^T J_i} \quad (13)$$

The above derivation matches the description in (24, 25): under certain conditions, a Bayesian inference problem reduces to an optimization problem, and the optimal solution can be found by an algorithm similar to expectation maximization. Specifically, the expectation term in (9) is an expectation step, and the updates in (13) are a maximization step. The values of the parameters are calculated based on the current values, and these values then used for the next iteration and so on until convergence. Because the independence of the estimated parameters is assumed, each parameter in $\mu$ (CBF and MTT) can be updated simultaneously to iteratively maximize the KL. Convergence is determined by observing the trend of changes in KL divergence between the approximate and true posteriors. When the change is less than a threshold, the iterating process terminated and the parameters at the current iteration serve as the final estimates.

Hence, the distribution $q(\mu_k) \sim N(m_k, \sigma_k^2)$ that approximates $q(\mu|Y)$ to $P(\mu|Y)$ can be found, where $\sigma_k$ contains the uncertainty information of the $k$th parameter estimated from a given scan. Instead of directly using the variance of each parameter, we use the coefficient of variation, which has the advantage of being independent of the intensity scale. It can simply be represented as a ratio between the standard deviation and the mean of a given parameter: $CoV = \sigma/m$, which is unitless. One can conclude that the higher the CoV, the more uncertainty there is associated with the parameter.

TABLE I Perfusion parameters for digital brain phantom chosen in a range of average values

|  | GRAY MATTER | WHITE MATTER | LESION |
|---|---|---|---|
| CBF (ML/100G/MIN) | 53 ± 14 | 25 ± 14 | 16 ± 4.25 |
| MTT (S) | 3.7 ± 0.7 | 4.6 ± 0.7 | 14 ± 0.75 |
| CBV (ML/100 G) | 3.3 ± 0.4 | 1.9 ± 0.9 | 3 ± 0.7 |

CBV- cerebral blood volume, CBF- cerebral blood flow, MTT- mean transit time.

TABLE II Scanner parameters in all simulations

| PARAMETER | VALUE |
|---|---|
| SCANNER MODEL | Siemens Definition |
| DETECTOR ROW NO. | 64 |
| FLYING FOCUS | On |
| NUMBER OF FRAMES | 50 |
| SCAN INTERVAL | 0 s, 1 s or 3 s |
| PITCH | 1.0 |
| ANGELS PER ROTATION | 600 |
| DETECTOR PIXEL SIZE | 0.5×1.0 mm |
| RECONSTRUCTED IMAGE SIZE | 512×512 |
| RECONSTRUCTED PIXEL SIZE | 1.0×1.0 mm |

## III. EXPERIMENTS
### A. Simulation studies

We simulated 3D (2D+t) CT perfusion scans using a modification of the brain phantom described in (26) (downloadable at https://www5.cs.fau.de/research/data/digital-brain-perfusion-phantom/index.html). We defined two annotated tissue classes to mimic a scan with ischemia: healthy tissue and tissue with reduced CBF and increased MTT (Fig. 2a). Perfusion parameters in Table I, in which was assigned according to (27), were assigned to the annotated tissues. From the perfusion phantom, we could generate dynamic projection data based on the forward model defined in (4) and (5). A total of 50 one-second frames were simulated, for which TACs are shown in Fig. 2b. Data were generated by forward projecting each frame with an acquisition protocol (Table II) based on a simulated scanner model. Poisson noise was added to the projections by setting the number of detected counts in a blank scan to 1e5. Noise-free and noisy projections were reconstructed to generate two sets of acquired dynamic images, from which perfusion parameters were inferred. In all simulations, reconstruction was performed with a helical Feldkamp–Davis–Kress algorithm (28,29). A total of four simulations were performed to evaluate the proposed method:

(1) *Comparison with conventional methods*

In the first simulation, we compared the proposed method with conventional methods with ideal simulated measurements. The noise-free dynamic images were used to compute the perfusion maps with both methods, and the bias was assessed by comparing image profiles across the lesion in perfusion maps generated from the SVD-based methods, the proposed method, and the ground truth. Note that only this simulation used the noise-free images, the rest using noisy simulated images.

(2) *Effect of scan interval*

In the second simulation, we evaluated scan protocols with different scan intervals using the proposed method. Here, scan interval indicates the time between the acquisition of the two frames. It was shown previously that scan interval variation could have an impact on perfusion imaging by introducing bias in the perfusion parameters (10,11). We varied the scan intervals by undersampling the simulated noisy dynamic frames. For a normal setup, the scan interval was zero; for long-interval setups, the scan intervals were either 1 s or 3 s, reducing the number of frames to 25 and 12, respectively. The hypothesis was that the Bayesian method can tolerate large intervals better than the SVD-based methods. Moreover, the measurement would be less reliable when fewer data were available, that is, the proposed Bayesian method would identify larger uncertainties in parameters. We also calculated the predicted enhanced TAC for a region-of-interest (ROI) in the lesion for Bayesian methods by applying the estimated distribution of the perfusion parameters to the forward model in (6). This enhanced TAC was then compared with the true one.

(3) *Effect of truncation*

In the third simulation, we evaluated the effect of truncated measurements on parameter inference with the proposed method. Truncation commonly occurs in CT perfusion when the acquisition is too short to capture the complete attenuation change due to poor cardiac output or large vessel occlusion (30). It has been shown that incomplete acquisition falsely reduces perfusion estimates and hence may alter patient management (12,13). To simulate such an effect, we made the last 20 or 30 s of the measurement unavailable for parameter inference. We hypothesized that the proposed Bayesian method would handle missing data better than the SVD-based method. Moreover, we hypothesized that the measurement would be less reliable because of the missing data, that is, the proposed Bayesian method would identify larger uncertainties in parameters. Again, we calculated the predicted enhanced TAC for an ROI in the lesion for the Bayesian method and compared it to the true one.

(4) *Effect of patient motion*

In the fourth simulation, we evaluated the effect of motion on parameter inference with the proposed method. Involuntary patient motion can create inconsistencies in temporal measurements and hence have a negative impact on perfusion map estimation (31,32). We examined the effect of motion on the uncertainty of the parameter estimates.

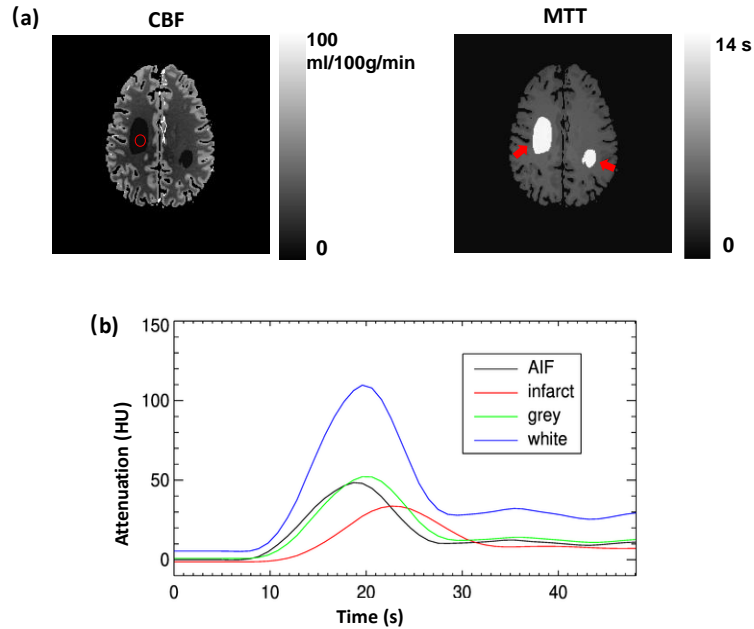

CBF-cerebral blood function, MTT-mean transit time, AIF-arterial input function

Fig. 2 Simulation phantom setup. (a) CBF and MTT perfusion phantoms from which data were generated. Two lesions with reduced CBF and increased MTT are labeled with arrows. (b) Time attenuation curves of voxels in different regions. AIF is scaled by 0.1 for visual inspection. A region-of-interest for quantification was drawn inside the lesion as indicated by the red circle.

For this investigation, we generated simulated motion-corrupted projections of the phantom by applying previously measured human motion (33) during forward projection and reconstructing the motion-corrupted images. Perfusion maps derived from the motion-free and motion-corrupted dynamic images were compared. The hypothesis was that the measurement would be less reliable in the motion case because of inconsistent data, that is, the proposed Bayesian method would identify larger uncertainties in parameters estimated from motion-corrupted data.

B. Patient study
The anonymized raw data of two patients who had previously undergone a head CT scan in the Department of Radiology at Westmead Hospital in Sydney, Australia, were collected with the approval of the Human Research Ethics Committee of the Western Sydney Local Health District. The written informed consent was obtained from the patients before each scan. The scans were acquired with a Definition 64 scanner

(Siemens Healthineers, Forcheim, Germany). Before each scan, 45 ml of iodine contrast agent was injected intravenously at a rate of 6 ml/s, followed by a 40 ml saline flush. After that, a total of 33 frames (26 × 1.5 s, 7 × 3 s) were acquired in axial (shuttle) mode. The first acquired frame was labeled as the baseline frame (mask). The reconstructions were initially performed with the vendor's filtered back-projection algorithm with a smoothing kernel of H40f. The pixel dimensions of the reconstructed image were 0.4 × 0.4 × 5 $mm^3$. DICOM volumes were exported without further post-processing.

One patient dataset was motion-free and the other presented visible motion artifacts. For the former, the proposed Bayesian method was directly applied to derive the perfusion parameters, while, for the dataset contaminated by the motion, we identified both inter- and intra-frame motion artifacts in the exported dynamic reconstructed frames. These images were labeled as **NMC**. Patient motion would be expected to result in greater uncertainty of parameter estimates, whereas a reduction in motion artifacts and blurring following the application of a motion correction algorithm would be expected to recover the reliability of the parameter estimates. As a test of the proposed method, we applied it to evaluate the effect of motion correction methods in terms of the perfusion parameters. Two motion correction methods were implemented in this study. The first method only compensated for inter-frame motion by directly operating on the exported images (34,35). All dynamic frames were rigidly registered to the first (baseline) frame to remove the inter-frame motion. The corrected images were labelled as **MC1**. The second method included an additional step before inter-frame motion compensation, which compensated for intra-frame motion using the method described in (36)(37). This iterative data-driven method jointly estimated the six degrees-of-freedom rigid motion of the head at each projection and the motion-corrected reconstruction from the raw projection data. At each iteration, head pose estimates were updated analytically at multiple projection angles by adjusting the head pose to reduce the differences between the forward and measured projections. The motion-corrected image was updated by taking the estimated motion into account during reconstruction with a fully 3D maximum likelihood expectation maximization or Feldkamp–Davis–Kress algorithm. Motion correction was applied to each frame independently. This was followed by inter-frame motion correction as in the first correction method. The corrected images were labeled as **MC2**. After motion correction, perfusion analysis was performed by applying the proposed Bayesian inference algorithms to use NMC, MC1, and MC2 images.

In contrast to the simulation study, AIF was unknown here and was derived directly from the images as follows. We first applied a threshold to the summation of all frames and drew a 2D ROI over the internal carotid region. We then corrected the AIF for the partial volume effect due to the small diameter of the target arteries—a venous output function was derived similarly in a straight sinus region, and the AIF was scaled to have the same area under the curve as the venous output function to account for the underestimation (38). After inference, hematocrit correction was performed on CBV and CBF (multiplying with a constant 0.733 (15)) to account for the different hematocrit values between arteries and capillaries.

## C. Quantification

All generated perfusion parameter images were compared with the truth in Fig. 2a by calculating the mean squared error and the bias. Denoting IM0 as the true image and IM1 as the image for comparison, we calculated the mean squared error MSE:

$$\text{MSE} = \frac{1}{N}\sum_{j}^{N}(\text{IM1}_j - \text{IM0}_j)^2,$$

where $N$ is the total number of voxels, and $j$ is the index of the voxel. The bias between two images was represented as:

$$\text{bias}(\%) = \left|\frac{\text{mean}(\text{IM1}_{\text{ROI}}) - \text{mean}(\text{IM0}_{\text{ROI}})}{\text{mean}(\text{IM0}_{\text{ROI}})}\right| \times 100\%.$$

This calculation was only performed in an ROI drawn inside the lesion, as shown in Fig. 2a. The CBF and MTT intensity distributions were also calculated for that ROI, which were also Gaussian distributed. The mean of the distribution was the mean at all voxels in the ROI; the standard deviation of the distribution was the mean of the square root of the variance at all voxels in the ROI. The wider the distribution, the greater the uncertainty of the estimated parameters in the ROI.

## IV. RESULTS
### A. Simulation studies
*(1) Comparison with conventional methods*
Fig. 3a shows the accuracy (MSE and bias) of CBF and MTT maps derived with the proposed method and with the deconvolution methods. Bias was calculated from the voxels of an ROI within a lesion (Fig. 2a). Fig. 3b shows profiles from the MTT images in the lower row of Fig. 3a, at the location indicated by the red dotted line. Compared with the SVD-based method, the proposed Bayesian method resulted in lower MSE and bias for CBF and MTT, and, therefore also, lower bias for CBV. The MTT image profile obtained with the Bayesian method corresponded well to the reference profile.

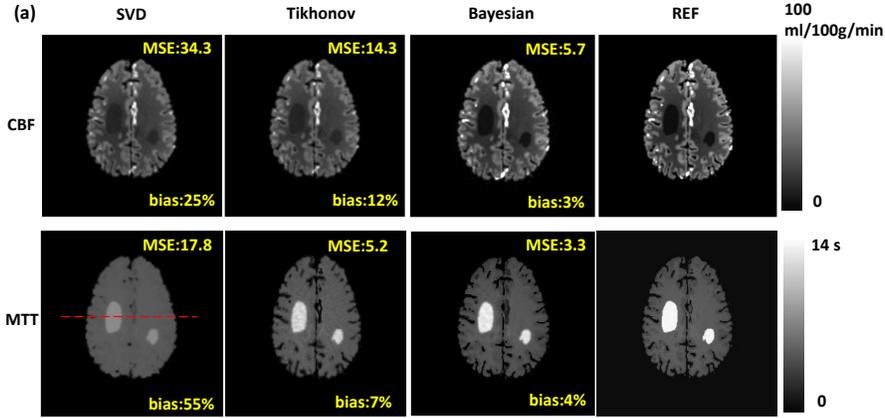

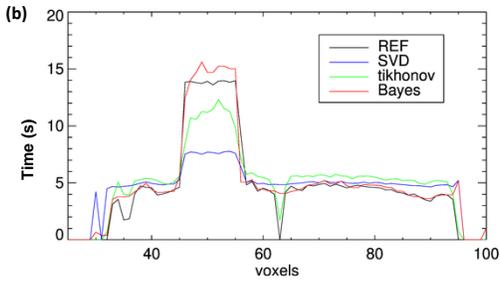

SVD-singular value decomposition, REF-reference

Fig. 3 A comparison of the results with SVD and the proposed method without noise in Simulation 1. (a) Compared with the true images (REF), the proposed method shows markedly less bias in lesions than the SVD-based methods for both CBF and MTT. (b) Profile plots (red dotted line in a) of MTT images show the reduction in bias with the proposed method.

(2) Effect of scan interval

In the second simulation, we assessed the impact of varying the scan interval on the MSE and bias of parameter estimates. As shown in Fig. 4a, the SVD-based methods resulted in markedly increased MSE and bias at large scan intervals, while the proposed method was much less affected by increased scan interval. At larger scan intervals, we observed that the mean CBF and MTT maps were noisier than those obtained with normal sampling (Fig. 4). The CoV maps and the CBF distribution plots indicated differences in CBF mean and variance with different scan intervals. The number of iterations used for full data, and scan intervals of 1 s and 3s were 6, 8 and 12, respectively. We also calculated the predicted enhanced TAC sampled in a lesion ROI for all methods in Fig. 5a.

(3) Effect of truncation

In the third simulation, our observations on the effect of truncation in Fig. 6 were similar to those regarding scan interval in Fig. 4. MSE and bias in MTT estimates were markedly degraded as truncation was increased with the SVD-based method while the proposed method was relatively unaffected. With severe truncation, the MTT estimate was biased and its variance higher, as indicated by the CoV images, compared to the case with full measurement data. The number of iterations of the Bayesian method for the full data and truncations of last 25s and 30s of data were 6, 6 and 9, respectively. Fig. 5b shows the predicted enhanced TAC sampled in a lesion ROI for all methods.

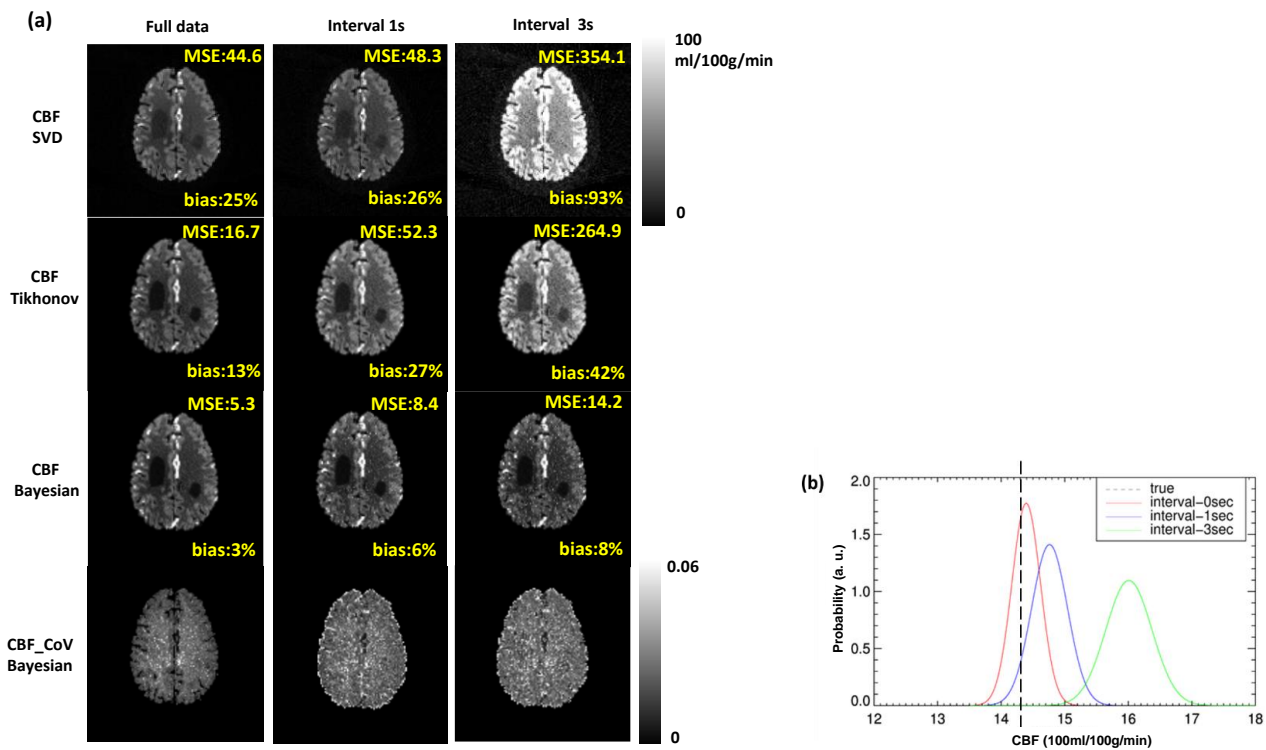

SNR-signal to noise, CBF-cerebral blood flow, CoV-coefficient of variation, SVD-singular value decomposition, MTT-mean transit time

Fig. 4 The effect of varying the scan interval in Simulation 2. For the full data, the SNR of the estimated CBF map (from the proposed method) was 3.2. (a) Mean CBF estimate and its CoV for three different scan intervals. The CBF and CoV values calculated with the SVD-based method have larger bias than those calculated with the Bayesian method, and bias increases with the interval. (b) The CBF intensity distribution of a sampled ROI (yellow circle in a) with different intervals. The wider the distribution, the greater the uncertainty in the parameters. Similar results were obtained for MTT (not shown).

*(4) Effect of patient motion*

In the fourth simulation with simulated motion, we observed that the mean CBF and MTT map had noticeable visual artifacts (Fig. 7a). The CBF_CoV images indicated that motion made the scan less reliable. High CoV values corresponded to regions with artifacts in the mean estimate maps, and CoV values were generally lower in the motion-free image. This suggests that CoV mapping is useful for identifying parts of the brain where parameter estimates are problematic due to motion-induced uncertainty. The numbers of iterations of the Bayesian method used for MC and NMC images were 6 and 7, respectively.

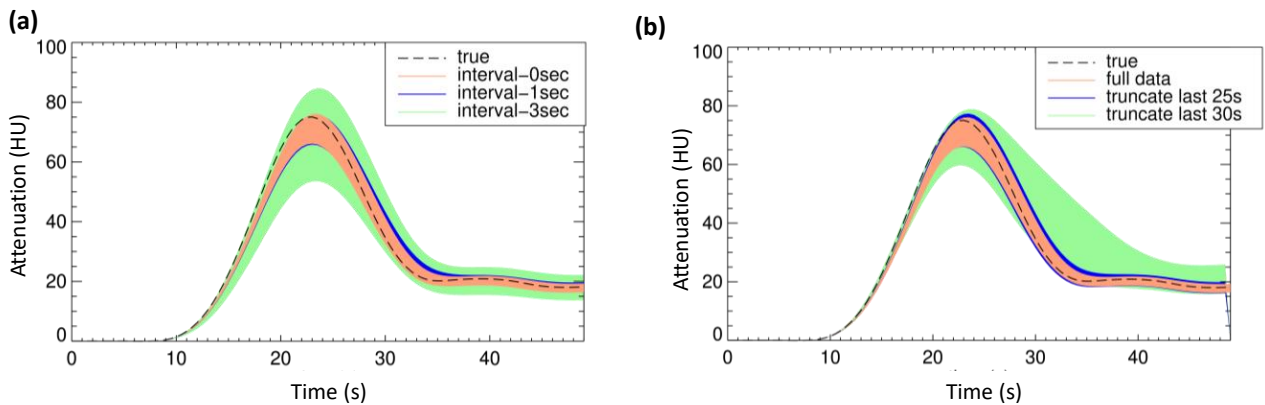

HU-Hounsfield unit

Fig. 5 Time-attenuation curves for the lesion region-of-interest from (a) simulation 2 and (b) simulation 3. With the inferred Gaussian distributed CBF and MTT (mean±SD), we can generate the lower and upper bounds of the time-attenuation curves. The predicted enhanced TAC with the proposed method and true enhanced TAC are well correlated. With more deterioration in data, the predicted enhanced TAC is more biased and with more uncertainty.

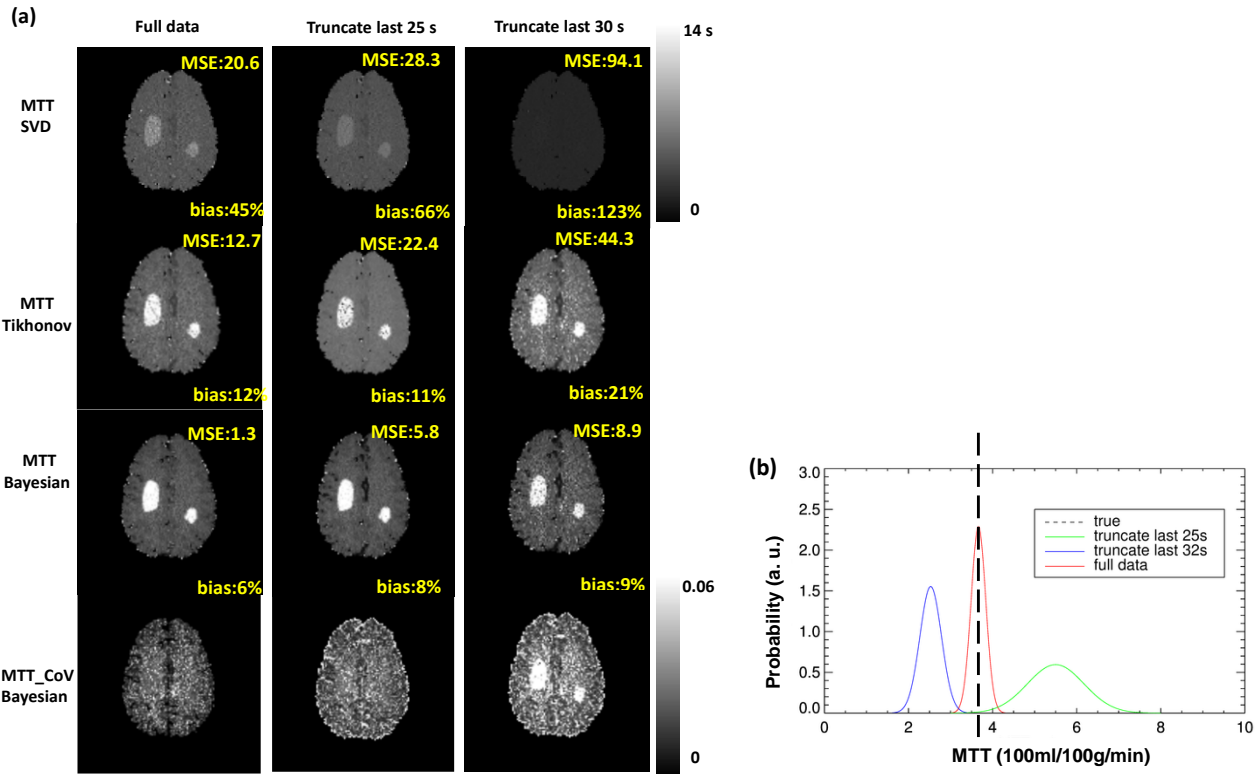

SNR-signal to noise, CBF-cerebral blood flow, CoV-coefficient of variation, SVD-singular value decomposition, MTT-mean transit time

Fig. 6 The effect of truncating the measured data in Simulation 3. For the full data, the SNR of the estimated MTT map (from the proposed method) was 2.8. (a) Mean MTT estimate and its CoV for full data and two different degrees of truncation. The MTT and CoV values calculated with the SVD-based method have larger bias, and the bias increases with the truncation. (b) The MTT intensity distribution of a sampled white matter region with different amounts of missing data. Similar results were obtained for CBF (not shown).

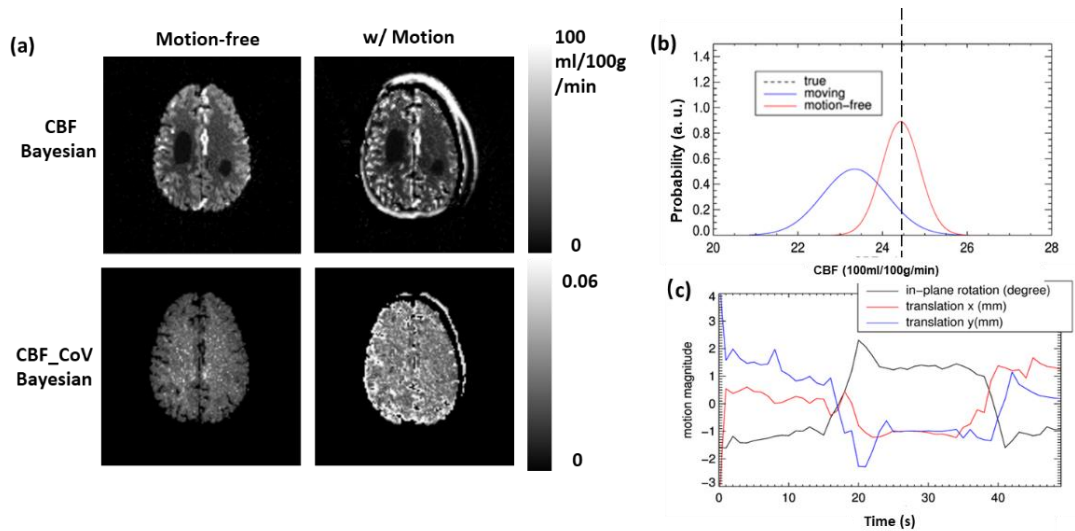

CBF-cerebral blood flow, CoV-coefficient of variation

Fig. 7 The effect of motion in Simulation 4. (a) Mean CBF estimate and CoV with and without simulated motion. Severe motion creates increased uncertainty in CBF estimates, as revealed in the CoV map. (b) The CBF intensity distribution of a sampled white matter region.  Similar results were obtained for MTT (not shown). (c) A segment of realistic rigid motion that is applied. Note that only in-plane motion is simulated (one rotation, two translations) in this 2D+t simulation.

B. Patient study
For the motion-free dataset, the proposed Bayesian method was directly applied to derive the perfusion parameters. According to the clinical report, this patient had an ischemic lesion due to a left middle cerebral artery infarct with increased MTT and a corresponding drop in CBF. Fig. 8 shows the CBF, MTT and CBV maps inferred using our method. The lesion and the normal tissue had clearly different hemodynamic parameters, indicated by the arrows in Fig. 8. The average CBF value within a region in the lesion was 13.2±4.1ml/100g/min, while for a region in the normal tissue it was 22.6±6.2 ml/100g/min. The average MTT value in the lesion was 22.3±4.8 s, while in the normal tissue it was 11.4±3.4 s. The average CBV values in the lesion and normal tissues were comparable. In addition, the uncertainty of the estimates can be assessed alongside the mean estimates as indicated in the middle column of Fig. 8.

For the dataset with motion artifacts, MC1 and MC2 were generated as described in the Methods section. The goal was to apply the proposed method to evaluate the effectiveness of two motion correction methods in terms of recovering the perfusion parameters. For both CBF and MTT, mean estimates and CoV were derived with MC1, MC2, and noMC. In Fig. 9a severe artifacts are visible with noMC. With MC1, the quality of the dynamic scan was successfully recovered, and the inferred parameters

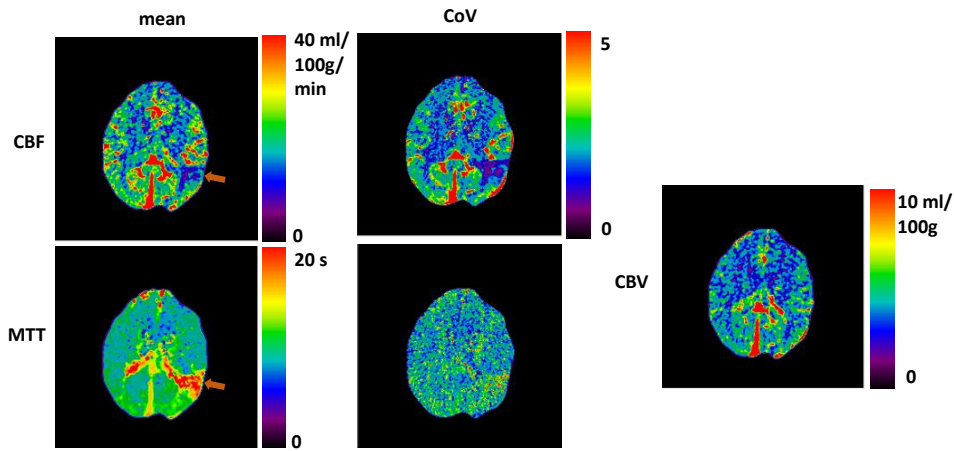

CBF-cerebral blood flow, MTT-mean transit time, CBV-cerebral blood volume, CoV-coefficient-of-variation

**Fig. 8** Mean and CoV maps inferred from a patient scan using the proposed method, in which a lesion due to the left middle cerebral infarct was indicated by the arrow. The average CBF value in the lesion was over 40 % lower than the normal tissue and the average MTT value in the lesion was over 96 % higher than the normal tissue. Nine iterations of the Bayesian method were performed.

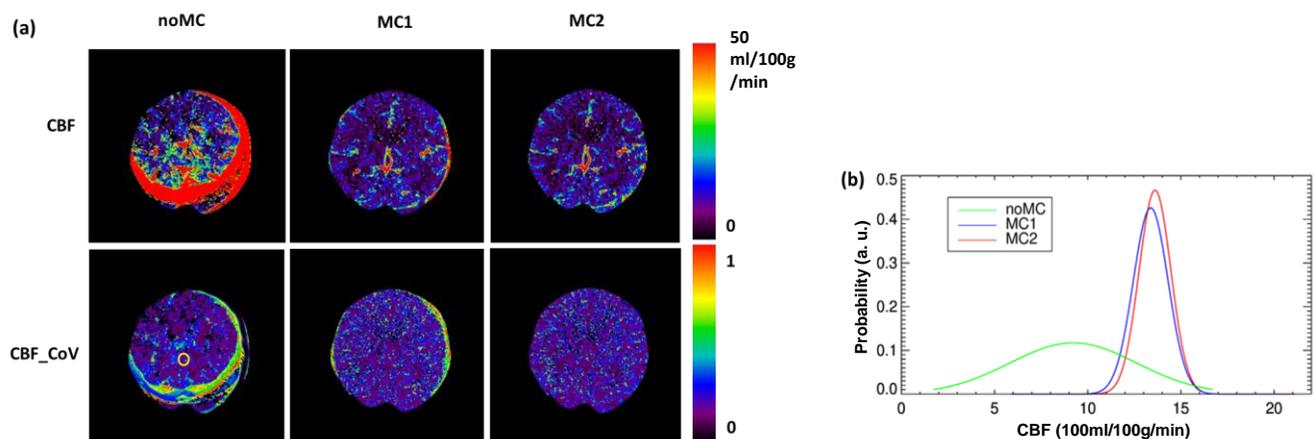

CBF-cerebral blood flow, CoV-coefficient of variation, noMC- non-corrected images, MC1- inter-frame corrected images, MC2- inter- and intra-frame corrected images were more reliable than those obtained with noMC, although there were still some residual artifacts. With MC2, the images were further improved, and the reliability of the estimation was the highest among the three. The results correspond well to the hypothesis of progressive recovery of the correction methods (from noMC to MC1 to MC2).

**Fig. 9** Mean and CoV maps of CBF inferred from a motion-corrected clinical CT perfusion scan using the proposed method. (a) MC2 yields fewer visible artifacts and lower overall CoV than MC1 or noMC. (b) CBF distributions of a selected ROI (yellow circle in a). MC2 provided the narrowest distribution, indicating that it provides the lowest uncertainty. Six, eight and twelve iterations of the Bayesian method were used for noMC, MC1 and MC2, respectively. Similar results were obtained for MTT (not shown).

## V. DISCUSSION

CT perfusion imaging results can be affected by many factors in practice, including the selection of scanner acquisition protocols, the choice of the post-processing technique, data truncation, and data perturbance caused, for example, by patient movement. All may affect the reliability of hemodynamic parameter estimation, which is crucial for accurate the diagnosis and appropriate treatment of stroke patients. Conventional inference methods cannot tolerate such imperfection well and cannot assess the reliability of the data, although such information could be useful diagnostically. We proposed a Bayesian method that can calculate CT perfusion maps and estimate the uncertainty at each voxel. The proposed method assumes better residual function hence the underlying perfusion model. Compared with conventional SVD-based methods, this approach has the advantages of (1) estimating parameters with less bias than conventional methods, especially when the measurements are imperfect and (2) providing the probability of an estimate, which can be used to guide the design of scan protocols or to compare the performance of different data processing or artifact reduction methods. We applied the proposed method to three simulated abnormal scan scenarios involving increased frame intervals, truncated measurement, and patient movement. The results indicated that the uncertainty measure provided by the method could allow scan protocols of varying reliability to be differentiated. In one patient study, the estimate CBF and MTT maps indicated an ischemic lesion. In a second patient study, we evaluated the proposed method on a stroke scan in which the patient exhibited severe motion. Two motion correction techniques were compared in terms of how much they improved the scan reliability, as assessed by the proposed Bayesian method.

The running time of the proposed method was longer than the conventional SVD-based methods, as Bayesian inference is an iterative process. For example, to produce the results in Figure 3, SVD took 2 s per slice while Tikhonov-SVD took 6 s and the Bayesian method took 10 s. Although the running time was longer for the Bayesian method, more accurate estimates and additional information about the distribution of the parameters of interest were obtained.

The proposed method successfully provided mean and uncertainty estimates for CBF and MTT. However, it is also possible to infer other parameters, such as bolus arrival time delay and permeability, if the associated acquisition model is assumed. For example, a delay exists between the injection time and the contrast agent arrival time in the tissue of interest, which has been shown to affect the CBF estimates, especially for negative time shifts and at high SNR (17). To include this effect, the convolution model in (1) could be modified, and the inference would proceed with minor change in future work. Permeability is an important indicator of the degree of blood-brain barrier damage. The Patlak model is commonly used to infer this parameter and could be incorporated with the proposed Bayesian method. However, a longer scan time (~200 s) is usually needed to provide enough information for the Patlak model, and such a scan protocol is not available at this time in our center.

Another possible way to infer the distribution of perfusion parameters is the frequentist method, which often requires repeatable measurements (which is not feasible for patients) or simulated measurements using techniques such as bootstrapping. We chose a Bayesian method in this study instead since we believe it is more appropriate for our application where the measured data are fixed and parameters are random. In addition, the Bayesian method has the advantage that prior knowledge can easily be included in the Bayesian inference process, whereas it is not obvious how to do this for bootstrapping (39). In the future work, spatial-temporal regularization could be included in the proposed method to reduce speckle noise in estimation and enable CT perfusion scans to be performed with lower tube current and radiation dose. A Variational Bayesian approach was chosen in this study. As mentioned earlier in the Introduction, MCMC is another popular Bayesian method shown to be efficient in inferring kinetic parameters in dynamic contrast-enhanced cardiac MR applications (40). MCMC can also incorporate the spatial prior information to reduce the speckle noise in estimation map (41).

We inferred the parameter distributions from dynamic reconstructed images. In principle, it is possible to infer the desired parameters directly from the measured raw projections without reconstruction. Applications of direct kinetic parameters estimation exist in CT (42), PET (43,44), and MRI (40,45,46). However, these approaches can only provide maximum likelihood estimates of parameters, not the full distribution. Another concern in CT perfusion is that it is difficult to separate the coupling between the attenuation and kinetic distribution since a perfect forward model can often not be guaranteed. For example, if there is patient motion during a scan, simultaneous motion correction and perfusion map inference can be problematic because errors from one side will likely propagate to the other.

A limitation of our measurement model is that the observation noise is assumed to be an independently and identically distributed (i.i.d.) Gaussian variable. Within each frame, the noise level is scaled according to the duration of that frame. This assumption leads to great convenience for computation. If the CT scan dose is very low, the actual noise model can be complicated (47,48) and the one assumed here may yield poor inference results. Other inference techniques may be required to consider the correct noise model. For example, techniques approximating the Poisson to Gaussian distribution may be required before applying Bayesian inference. Another limitation within the proposed measurement model is the bolus arrival delay was not considered and hence inferred. As stated earlier, the model can be modified to account for this effect. A further clinically relevant limitation is that we did not evaluate the estimated parameter maps for identifying the infarct core and penumbra region, which could be relevant to a clinical decision. A more detailed investigation of the clinical impact of the method is needed.

## VI. CONCLUSION

We proposed and applied a Bayesian approach that infers the hemodynamic parameters in CT perfusion imaging. The method can yield mean parameter estimates and uncertainty estimates of such parameters under imperfect scan scenarios. It can be used to assess the confidence of parameter estimates under imperfect scan conditions to evaluate protocol designs. Further studies are required to fully evaluate the clinical utility if the proposed method.

## Acknowledgement

Funding: This work was partly funded by Key Laboratory for Magnetic Resonance and Multimodality Imaging of Guangdong Province (2020B1212060051). The authors wish to thank Johan Nuyts from KU Leuven for the discussion; Krystal Moore from the Department of Radiology at Westmead Hospital for collecting the datasets; and Karl Stierstorfer, Siemens Healthcare, Forchheim, Germany, for his help with reading the raw CT data.

## Footnote

Conflict of interest: The authors have no conflict of interest to declare.

Ethical statement: The authors are accountable for all aspects of the work in ensuring that questions related to the accuracy or integrity of any part of the work are appropriately investigated and resolved. All human studies have obtained the approval of the Human Research Ethics Committee of the Western Sydney Local Health District. The written informed consent was obtained from the patients before each scan.

Figure legends

Fig. 1 The indicator dilution theory model (a) describes the blood supply to the tissue at one voxel in the brain. After contrast injection, the bolus enters the voxel via an arterial inlet Ca(t), is dilated in the capillary bed, and leaves the voxel via the venous outlet. Due to the limited spatial resolution of the CT image, Y(t) contain numerous capillary beds as well as arterioles and venules. (b) is the residual function h(t) that describes the hemodynamic functions in the physiological model. CBF, CBV, and MTT can be derived from h(t).

Fig. 2 Simulation phantom setup. (a) CBF and MTT perfusion phantoms from which data were generated. Two lesions with reduced CBF and increased MTT are labeled with arrows. (b) Time attenuation curves of voxels in different regions. AIF is scaled by 0.1 for visual inspection. A region-of-interest for quantification was drawn inside the lesion as indicated by the red circle.

Fig. 3 A comparison of the results with SVD and the proposed method without noise in Simulation 1. (a) Compared with the true images (REF), the proposed method shows markedly less bias in lesions than the SVD-based methods for both CBF and MTT. (b) Profile plots (red dotted line in a) of MTT images show the reduction in bias with the proposed method.

Fig. 4 The effect of varying the scan interval in Simulation 2. For the full data, the SNR of the estimated CBF map (from the proposed method) was 3.2. (a) Mean CBF estimate and its CoV for three different scan intervals. The CBF and CoV values calculated with the SVD-based method have larger bias than those calculated with the Bayesian method, and bias increases with the interval. (b) The CBF intensity distribution of a sampled ROI (yellow circle in a) with different intervals. The wider the distribution, the greater the uncertainty in the parameters. Similar results were obtained for MTT (not shown).

Fig. 5 Time-attenuation curves for the lesion ROI from (a) simulation 2 and (b) simulation 3. With the inferred Gaussian distributed CBF and MTT (mean±SD), we can generate the lower and upper bounds of the time-attenuation curves. The predicted enhanced TAC with the proposed method and true enhanced TAC are well correlated. With more deterioration in data, the predicted enhanced TAC is more biased and with more uncertainty.

Fig. 6 The effect of truncating the measured data in Simulation 3. For the full data, the SNR of the estimated MTT map (from the proposed method) was 2.8. (a) Mean MTT estimate and CoV for full data and two different degrees of truncation. The MTT and CoV values calculated with the SVD-based method have larger bias, and the bias increases with the truncation. (b) The MTT intensity distribution of a sampled white matter region with different amounts of missing data. Similar results were obtained for CBF (not shown).

Fig. 7 The effect of motion in Simulation 4. (a) Mean CBF estimate and CoV with and without simulated motion. Severe motion creates increased uncertainty in CBF estimates, as revealed in the CoV map. (b) The CBF intensity distribution of a sampled white matter region. Similar results were obtained for MTT (not shown).

Fig. 8 Mean and CoV maps inferred from a patient scan using the proposed method, in which a lesion due to the left middle cerebral infarct was indicated by the arrow. The average CBF value in the lesion was over 40 % lower than the normal tissue and the average MTT value in the lesion was over 96 % higher than the normal tissue. Nine iterations of the Bayesian method were performed.

Fig. 9 Mean and CoV maps of CBF inferred from a motion-corrected clinical CT perfusion scan using the proposed method. (a) MC2 yields fewer visible artifacts and lower overall CoV than MC1 or noMC. (b) CBF distributions of a selected ROI (yellow circle in a). MC2 provided the narrowest distribution, indicating that it provides the lowest uncertainty. Six, eight and twelve iterations of the Bayesian method were used for noMC, MC1 and MC2, respectively. Similar results were obtained for MTT (not shown).